\documentclass[submission,copyright,creativecommons]{eptcs}
\providecommand{\event}{SOS 2022} % Name of the event you are submitting to

\usepackage{iftex}

\ifpdf
  \usepackage{underscore}         % Only needed if you use pdflatex.
  \usepackage[T1]{fontenc}        % Recommended with pdflatex
\else
  \usepackage{breakurl}           % Not needed if you use pdflatex only.
\fi

\usepackage{amsmath,amssymb}
\usepackage{xcolor}
\usepackage{bbm}
\usepackage{stmaryrd}
\usepackage{mathbbol}
\usepackage{mathtools}
\usepackage{upgreek}
\usepackage{url}
\usepackage{fancyvrb}
\usepackage[mathscr]{euscript}

\usepackage{float}
\usepackage{listings}

\newfloat{lstfloat}{htbp}{lop}
\floatname{lstfloat}{Listing}

\newcommand{\comma}{\mathrel{\mbox{\tt ,}}}

\newcommand{\lolli}{\multimap}
\newcommand{\send}{\rightarrow}
\newcommand{\eventD}{\gg}
\newcommand{\tostate}{\Rightarrow}
\newcommand{\Time}{\mathbb{t}}

\newcommand{\smallang}{\mbox{{\sf \emph{Stipula}}}}
\newcommand{\Stipula}{\mbox{{\sf \emph{Stipula}}}}

\newcommand{\contract}{\mathbb{C}}

\newcommand{\zero}{\mbox{\raisebox{-.6ex}{{\tt -\!\!\!-}}}}

\newcommand{\lred}[1]{\mathrel{\stackrel{#1}{\longrightarrow}}}

\newcommand{\xred}[1]
{ \setbox0=\hbox{$\, {}^{#1}\, $}
  \setbox1=\hbox{$\longrightarrow$}
  \loop\setbox1=\hbox{$-$\kern-0.3em\unhbox1}\ifdim\wd1<\wd0\repeat
  \hbox{$\ \mathop{\box1}\limits^{#1} \ $}
}

\newcommand{\wt}[1]{\vect{#1}}

\newcommand{\vect}[1]{\overline{#1}}

\title{From Legal Contracts to Legal Calculi:\\ the code-driven normativity}
\author{Silvia Crafa
\institute{Dipartimento di Matematica "Tullio Levi-Civita"\\ Universit\`a di Padova\\ Italy}
\email{silvia.crafa@unipd.it}
}
\def\titlerunning{From Legal Contracts to Legal Calculi}
\def\authorrunning{S. Crafa}
\begin{document}
\maketitle

\begin{abstract}
%There is an increasing trend, called Code-Driven Law, 
Using dedicated software to represent or enact legislation or regulation has the advantage of solving the inherent ambiguity of legal texts and enabling the automation of compliance with legal norms. On the other hand, the so-called code-driven normativity is less flexible than the legal provisions it claims to implement, and transforms the nature of legal protection, potentially reducing the capability of individual human beings to invoke legal remedies.

In this article we focus on software-based legal contracts;
%. We put forward a number of remarks about the suitable abstraction level for legal programming languages, 
we illustrate the design of a legal calculus  whose primitives allow a direct formalisation of contracts' normative elements (\emph{i.e.}, permissions, prohibitions, obligations, asset transfer, judicial enforcement and openness to the external context). We show that interpreting legal contracts as interaction protocols between (untrusted) parties enables the generalisation of formal methods and tools for concurrent systems to the legal setting.
\end{abstract}

\section{Code is law, really?}

Ethereum's smart contracts popularised the \emph{Code is Law} principle\footnote{originally proposed by Lawrence Lessing \cite{Lessig99}}, that is the idea of relying on software code to provide unambiguous %and transparent 
definition and automatic execution of transactions between (mutually untrusted) parties; and  when in disputes, the code of the contract, which is always publicly available, shall prevail. This principle is  rooted in the blockchain's dogma that trust is hardwired into intermediary transparent algorithms.
On this account, several governments have recognised that smart contracts, and more generally programs operating over distributed ledgers, may indeed have legal value~\cite{WyomingAct,ItalianLaw2019,MaltaAct}. 

This approach encompasses the blockchain technologies, since most of the benefits of digitally encoding legally binding agreements come from the precise definition and the automatic execution of a piece of programmable software, not necessarily operating over a blockchain. Accordingly, there is an increasing trend, called \emph{Code-Driven Law} \cite{Cohubicle}, using dedicated software to represent or enact legislation or regulation. Technologies like Rules as Code~\cite{RulesAsCode}, Catala~\cite{CatalaLang} or Akoma Ntoso~\cite{AkomaNtoso} propose to create a machine-consumable version of some types of rules issued by governments and public administrations, \emph{e.g.}, the tax office, student grant provision or social security agency. 
This helps identify potential inconsistencies in regulation, reduce the complexity and the ambiguity of legal texts and support the automation of legal decisions by the code-driven enforcement of rules:
instead of relying on \emph{ex-post} enforcement by third parties (\emph{i.e.}, courts and police), the rules hardwired into code are enforced \emph{ex-ante}, making it very difficult for people to breach them in the first place \cite{FilippiHassan2016}. 

However, transposing legal rules into technical rules is a delicate process, since the inherent ambiguity of the legal system is necessary to ensure a proper application of the law on a case-by-case basis. Regulation by code is instead always more specific and less flexible than the legal provisions it purports to implement, thereby giving software developers and engineers the power to embed their own interpretation of the law into the technical artefacts that they create \cite{FilippiHassan2016}. More precisely, the process of translating parties' intentions, promises, actions, powers and prohibitions into computer code, although public and unambiguous for the machine, is problematic and does not solve the problem but moves it into another dimension (\cite{FormeFalso}). 
Secondly, the code-driven law is based on the the automation of compliance with pre-set rules: if certain conditions are met the code will self-execute whatever it was programmed to do, not leaving room for disagreement about the right way to interpret the norms. Even if the need for judicial arbitration cannot be eliminated (\emph{e.g.} one always has the right to appeal to the court if the code adopted an incorrect tax rate), 
the code-driven normativity transforms the nature of legal protection potentially reducing the capability of individual human beings to invoke legal remedies \cite{Cohubicle}.

As an example, the Ethereum's code-is-law dogma declined with the famous TheDAO attack \cite{TheDAO}. Indeed, from the code-is-law perspective, a problem in the source code leading to unexpected behaviour of the smart contract, is a feature of the code and not an error. But the first hard fork of the Ethereum blockchain showed that this principle is not satisfactory in practice: when large volumes of money are at stake, no one is really willing to consider a security error in a program as part of the contract they have signed. Moreover, a less naive look nowadays leads us to state that blockchain does not hardwire trust into algorithms, but rather reassigns trust to a whole series of actors (miners, programmers, companies and foundations) who implement, manage and enable the functioning of this technological platform.

%\section{Legal Contracts as interaction protocols in concurrent systems} % as concurrent software
\section{Form Legal Contracts to Legal Calculi}

Despite the difficulties highlighted above, a sensible process of digitisation of legal texts has clear advantages. In this article  we discuss a specific line of research, conducted in collaboration with Cosimo Laneve and Giovanni Sartor, focusing on a specific subset of legal documents, that is the legal contracts (\cite{techReport,ProLaLa22,Festschrift} and other submitted articles). Legal contracts are defined as “those agreements that are intended to give rise to a binding legal relationship or to have some other legal effect” \cite{CommonFramereference2009Pr}. The principle of \emph{freedom of form} in contracts, which is shared by modern legal systems, says that parties are free to express their agreement using the language and medium they prefer, including a programming language. Therefore, by this principle, software-based contracts may count as legal contracts.
However, a contract produces the intended effects, declared by the parties, only if it is legally valid: the law may deny validity to certain clauses (\emph{e.g.}, excessive interests rate) and/or may establish additional effects that were not stated by the parties (\emph{e.g.}, consumer’s power to withdraw from an online sale, warranties, etc.). 
%The intervention of the law is particularly significant when the contractor (usually the weaker party, such as the worker in an employment contract or the consumer in an online purchase) agrees without having awareness of all clauses in the contract, nor having the ability to negotiate them, due to the existing unbalance of power.
Moreover, the contract's institutional effects are guaranteed by the possibility of activating judicial enforcements. That is, each party may start a lawsuit if she believes that the other party has failed to comply with the contract. 
%In this case, the judge will have to interpret the contract, ascertain the facts of the case, and determine whether there has indeed been a contractual violation. Accordingly, the defaulting party may be enjoined to comply or pay damage
%
Therefore, the assimilation of software-based contracts to legally binding contracts, or rather the double nature of digital contracts as computational mechanisms and as legal contracts, raises both legal and technological issues. 

First of all, in \cite{techReport} we observe that different kinds of software-based solutions can be valuable in the different phases of the lifecycle of a legal contract, which goes through negotiation, contract storage/notarizing, performance, enforcement and monitoring, possible modification and dispute resolution. Accordingly, several projects are being developed for defining %human readable
code-driven legal contracts, \emph{e.g.}~\cite{Lexon,Accord,OpenLaw,SPESC,SLCML,Orlando,Catala}.
We focus here in the problem of defining suitable programming languages to write legal contracts, since 
finding the suitable abstraction level for legal languages is still an open issue. Indeed, such a language should be easy-to-use and to understand for legal practitioners, but at the same time, the language should be fairly expressive, have a running environment with a precise semantics, and possibly supply sensible analyzers.

The solution we discuss in this article is the \Stipula\ programming language, whose design is based on the following main remarks:
\begin{itemize}
\item it is an \emph{intermediate domain-specific language}: a core calculus more concrete than a user-friendly contract specification language, and more abstract than a full-fledged programming language. This is in line with the research approach of desugaring the high level programming language into a core \emph{Legal Calculus}~\cite{LegalCalculi,DwivediEtAl}, pivoted on few selected, concise and intelligible primitives, together with a precise formalisation. % influenced by principles of concurrent systems~\cite{MilnerBook}.
This is the case of the Catala~\cite{Catala}  language for modelling statutes and regulations clauses, the Orlando~\cite{Orlando} language for modelling conveyances in property law, and the Silica language~\cite{ObsidianPaper} language for generic smart contracts; 
\item the basic primitives of \Stipula\ has been designed to easily map the \emph{building blocks of legal contracts} into template programs and design patterns. Therefore, the direct formalisation of normative elements (\emph{i.e.}, permissions, prohibitions, obligations, judicial enforcement and openness to the external context) as programming patterns, increases the transparency and the understanding of the link between executable instructions and institutional-normative effects;
\item  a legal contract is interpreted as an \emph{interaction protocol}, that dynamically regulates permissions, prohibitions and obligations between parties, which behave concurrently as time flows. Accordingly, the definition of \Stipula\ is influenced by the theory of concurrent systems, both in the definition of the operational semantics (with a precise control of nondeterminism) and in the definition of a bisimulation-based observational equivalence, that equates contracts that are syntactically different but are legally equivalent since they exhibit the same observable normative elements;
\item the language definition is \emph{implementation-agnostic}, and can be either implemented as a centralised platform or it can be run on top of a distributed system, such as a blockchain.
Implementing \Stipula\ in terms of smart contracts (\emph{e.g.}, compiling in Solidity), would bring in the advantages of a public and decentralised blockchain platform. However, digital legal contracts are more general and encompass smart contracts: they can provide benefits in terms of automatic execution and enforcement of contractual conditions, traceability, and outcome certainty even without using a blockchain. In particular, running a legal contract over a secured centralised system allows for more efficiency, energy save, additional privacy. Moreover, a controlled level of intermediation can better monitor the contract enforcement, dealing with disputes between contract’s parties and carrying out judicial enforcements. 
%Finally, the intrinsic open nature of legal contracts is another challenge for smart contracts, that can hardly deal with the off-chain world: external data enter the blockchain only through oracles, which are problematic in many senses. 
A prototype centralised implementation of \Stipula\ as a Java application is available in \cite{stipulaprototype}. %, and the dynamic change of behaviours conflicts with the rigidity of smart contracts definition. Time is another big issue in blockchains.
%the distinctive language elements of Stipula are quite abstract, and only a concrete implementation of Stipula can properly address this issu
\end{itemize}
We think that, even if only a concrete implementation can properly address specific issues, studying the theory of a domain-specific legal calculus is a first interesting step, that sheds some light on the digitalisation of legal texts.

\section{Stipula and the code-driven normativity}

\begin{figure}[t]
\begin{center}
\begin{tabular}{|@{\quad}l@{\quad}|@{\quad}l@{\quad}|}
\hline
\qquad {\bf legal contracts} & \qquad {\Stipula} {\bf contracts}
\\[.2cm] \hline
meeting of the minds & agreement primitive
\\[.2cm] \hline
permissions, prohibitions & state-aware programming 
\\[.2cm] \hline
obligations & event primitive
\\[.2cm]
\hline
%\\
currency and tokens & asset-aware programming
\\[.2cm] \hline 
openness to the environment &  intermediary pattern 
\\[.2cm] \hline
judicial enforcement and exceptional behaviours  & authority pattern %party and ad-hoc pattern 
\\[.2cm] \hline
\end{tabular}
\end{center}
\caption{\label{fig:correspondence} Correspondence between legal elements and {\Stipula}
features}
\end{figure}

A preliminary interdisciplinary research recognised that most real legal contracts are written by combining the following basic elements:
\begin{enumerate}
\item the \emph{meeting of the minds}, that involves the contract’s subscribers to accept the terms of the contract, and identifies the moment when legal effects are triggered;
\item a number of \emph{permissions}, \emph{prohibitions} and \emph{obligation} clauses that may dynamically change, \emph{e.g.}, the permission to use a good until a deadline;
\item transfer of \emph{currency} or other \emph{assets}, \emph{e.g.}, the property of a physical or digital good, to be used for payments, escrows and securities;
\item the \emph{openness} to external conditions or data, \emph{e.g.}, a triggering condition depending on the value of a stock at a given date;
\item the possibility of activating \emph{judicial enforcements} triggered by a dispute resolution mechanism or by a third party monitoring conditions that can be hardly digitalised, as the diligent care or the good faith.
\end{enumerate}

Accordingly, the basic primitives of \Stipula\ has been designed to easily map these building blocks of legal contracts into template programs and design patterns, as summarised in Figure~\ref{fig:correspondence}. More precisely, the {\tt agreement} construct directly encodes the meeting of the minds. 
Normative elements are expressed by a strictly regimented behaviour in legal contracts: permissions and empowerments correspond to the possibility of performing an action at a certain stage, prohibitions correspond to the interdiction of doing an action, while obligations are recast into commitments that are checked at a specific time limit and
%(2) Permissions and prohibitions are represented by the possibility and the impossibility for a party to invoke a function at a certain contract’s stage.
% Obligations are enforced by scheduling an event that checks the fulfilment at a given date, issuing 
issue a corresponding penalty if the obligation has not been met. 
Moreover, to model the dynamic change of the set of normative elements according to the actions that have been done (or not), \Stipula\ commits to a \emph{state-aware programming style}, inspired by the state machine pattern widely used in smart contracts (c.f. Solidity \cite{SoliditySM} and Obsidian \cite{Obsidian}). This technique allows one to enforce the intended behaviour by prohibiting, for instance, the invocation of a function before another specific function is called.

In order to promote an \emph{asset-aware programming} (\cite{Flint,Nomos,Move}), assets are a specific value type, and asset manipulation is syntactically distinguished from standard operations, to stress the fact that assets cannot be destroyed nor forged but only transferred. Contract clauses depending on external data are implemented by means of a party that takes the role of intermediary and assumes the legal responsibility of timely retrieving data from the external source agreed in the terms of the contract (see the bet contract below). The contract's intermediary need not to be a third party authority, but one of the party can assume also the role of intermediary, provided that all the others agree. This is different from relying on Oracles web services, to whom legal responsibilities can hardly be attributed.
Finally, dispute resolutions, judicial enforcement of legal clauses and exceptional behaviours due, \emph{e.g.}, to force majeure, are implemented by including in the contract a party that %takes the role of authority with corresponding functionalities.
takes the legal responsibility of interfacing with a court or an Online Dispute Resolutions platform\footnote{as The European ODR platform at {\tt https://ec.europa.eu/consumers/odr.}}.

%\section{Programming Legal Contracts in Stipula }
%!TEX root = Stipula.tex

We illustrate the expressivity of {\smallang} by showing the contracts for a set of archetypal acts (taken form \cite{techReport}). They are simple but they represent the distinctive elements that can be found in most contracts. 

\subsection{Subscription contract: obligation of periodic payment}

We define  a simple contract representing the annual subscription to a magazine or a service.
Upon subscription the buyer must pay a deposit, then she must pay the annual fee. If she has not paid within one  month, the deposit is transferred to the editor. At the end of the year an event changes the status of the contract so to enable the payment of the annual fee with a maximum delay of one month. If the buyer is up to date with the payments, she can terminate the subscription and get back the deposit.
%Alla sottoscrizione deve versare una \texttt{cauzione}, poi deve pagare la fee annuale. Se entro 1 mese non ha pagato, la cauzione viene trasferita all'editore. Alla fine dell'anno un evento riporta lo stato da \texttt{Payed} a \texttt{To\_Pay}.Anche qui gli obblighi sono rappresentati da controlli ad un tempo limite (\texttt{1 month}).

\begin{figure}[t]
{\small
\begin{lstlisting}[numbers=left,,numberstyle=\tiny,mathescape,basicstyle=\ttfamily,caption={The subscription contract},captionpos=b,label=subscribe] 
stipula Subscription {
    assets wallet
    fields cost, deposit

    agreement (Editor,Buyer) {
        Editor, Buyer: cost, deposit 
    } $\tostate$ @Inactive

    @Inactive Buyer : subscribe [h] 
        (h == deposit) {
            h $\lolli$ wallet 
            now + 1 month $\eventD$ @To_Pay { wallet $\lolli$ Editor } $\tostate$ @End	
    } $\tostate$ @To_Pay

    @To_Pay Buyer : annualFee [h]
        (h == cost) {
            h $\lolli$ Editor  
            now + 1 year  $\eventD$ @Payed {} @To_Pay
            now + 1 year + 1 month $\eventD$ @To_Pay { wallet $\!\!\lolli\!\!$ Editor } $\tostate$ @End
    } $\tostate$ @Payed
    
    @Payed Buyer : terminate {
           wallet $\lolli$ Buyer
    }$\tostate$ @End
}
\end{lstlisting}
}
\end{figure}

The code in Listing~\ref{subscribe} shows that a contract is similar to a class in an OOL, containing a set of fields, a constructor and a number of functions. Contract's fields are distinguished into standard fields ({\tt cost} and {\tt deposit} store numbers corresponding to the fees and the deposit) and assets. The contract's asset field {\tt wallet} is initially empty and will hold the buyer's money in escrow.
The agreement (lines 5-7) is a sort of constructor for the contract: it is intended as a multiparty synchronization between the parties, \emph{i.e.} {\tt Editor} and {\tt Buyer}, who have to agree about the initial values of {\tt cost} and {\tt deposit}. After the agreement has been reached, the contract enters into the initial state {\tt @Inactive}.

The possible states of the contract are {\tt @Inactive, @To_Pay, @Payed}, and the contract's functions {\tt subscribe}, {\tt annualFee} and {\tt terminate} are defined so that only the buyer (who subscribed the agreement) can call them, and {\tt subscribe} can be called only once at the beginning. The parameter {\tt h} is an amount of assets, and a pre-condition checks that it corresponds to the expected amount. The operation {\tt h $\lolli$ wallet} transfers the assets {\tt h} into the constract's {\tt wallet}, while {\tt h $\lolli$ Editor} moves them to the editor.

Lines 12,18 and 19 issue the events corresponding to the annual payment obligation. Line 12 and 19, schedule an event that, after one month from now, resp. form the end of the paid year, check whether the (first) annual fee has not been paid (\emph{i.e.} the state is still {\tt To_Pay}), and in that case transfer the deposit to the editor and terminate the contract. Line 18 issues an event that in a year's time will allow the new payment by moving the contract's state from {\tt @Payed} back to {\tt @To_Pay}. Finally, the buyer is allowed to terminate the subscription only if all payments are regular; accordingly, the function {\tt terminate} can be invoked only in state {\tt @Payed} and the deposit is refunded to the buyer.

\subsection{The Digital Licensee contract: usage and purchase, dispute resolution}

%{\footnotesize
 \begin{lstfloat}
%\begin{figure}[t]
 {\small
\begin{lstlisting}[numbers=left,numberstyle=\tiny,mathescape,basicstyle=\ttfamily,caption={The contract for a digital licence},captionpos=b,label=DigitalLicence] 
stipula Licence {
  assets token, wallet  
  fields cost, t_start, t_limit 
 
  agreement (Licensor,Licensee,Authority){  
    Licensor, Licensee : cost, t_start, t_limit 
   } $\tostate$ @Inactive

  @Inactive Licensor : offerLicence [t] { 
    t $\lolli$ token 
    now + t_start $\eventD$ @Proposal {  token $\lolli$ Licensor } $\tostate$ @End
  } $\tostate$ @Proposal

  @Proposal Licensee : activateLicence [h]
    (h == cost){
       h $\lolli$ wallet
       wallet*0,1 $\lolli$  wallet, Authority 
       uses(token,Licensee) $\send$ Licensee 
       now + t_limit $\eventD$  @Trial {
                      wallet $\lolli$ Licensee 
                      token $\lolli$  Licensor 
                    } $\tostate$ @End
  } $\tostate$ @Trial

  @Trial Licensee : buy {
    wallet $\lolli$ Licensor 
    token $\lolli$ Licensee 
  } $\tostate$ @End

  @Trial Authority : compensateLicensor {
    wallet $\lolli$  Licensor 
    token $\lolli$  Licensor 
  } $\tostate$@End

  @Trial Authority : compensateLicensee {
    wallet $\lolli$ Licensee 
    token $\lolli$ Licensor; 
  } $\tostate$ @End
}
\end{lstlisting}
}
\end{lstfloat}
%\end{figure}
%}

Let us consider a contract corresponding to a licence to access a digital service, like a software or an ebook: % inspired form a similar example in \cite{Sartor2018},\Silvia{controlla se riusciamo a codificare la versione originale} 
the digital service can be freely accessed for a while, and can be permanently bought with an explicit communication within the evaluation period (for a similar example, see \cite{Sartor2018}). The licensing contractual clauses  can be described as follows:
%\Silvia{sono clausole scritte legalmente corrette? soprattutto Art. 3 e 4. GIOVANNI: Mi Ho 
%aggiunto un pezzettino alla 4. Mi sembra che la 3 non sia catturata dalla nostra 
%formalizzazione. Io la cancellerei. SILVIA: l'art. 3 serve per giustificare l'Authority, 
%spiegato nel testo} 
\begin{description}
\item[Article 1.]
{\tt Licensor} grants {\tt Licensee} for a licence to evaluate the product
 and fixes (\emph{i}) the
\emph{evaluation period} and (\emph{ii}) the \emph{cost} of the product if 
{\tt Licensee} will 
bought it.
\item[Article 2.] 
{\tt Licensee} will pay the product in advance; he will be reimbursed if the product 
will not be bought with an explicit communication within the evaluation period.
 The refund will be the 90\% of the cost
because the 10\% is payed to the {\tt Authority} (see Article 3).
\item[Article 3.] 
{\tt Licensee} must not publish the results of the evaluation during the evaluation period
and {\tt Licensor} must reply within 10 hours to the queries of {\tt Licensee} related to the 
product; this is supervised by {\tt Authority} that may interrupt the licence and reimburse
either {\tt Licensor} or {\tt Licensee} 
according to whom breaches this agreement.
\item[Article 4.]
This license will terminate automatically at the end of the evaluation period, if the {\tt Licensee} does not buy the product.
\end{description}

Compared to the previous example, 
%this contract involves payment and refund: 
%an amount of currency is escrowed, and two parts of it will be sent to different parties, the {\tt Authority} and either the {\tt Licensor} or the {\tt Licensee}. {\smallang} provides the general \texttt{asset} abstraction, together with a general operation to move just a (positive) subset of the asset to a different owner. 
%This is exactly what is needed to deal with currency, therefore the {\smallang} 
the licence contract holds two different assets: an indivisible non fungible \texttt{token} providing an handle to the digital service, and a \texttt{wallet} that is a fungible asset corresponding to the amount of currency kept in custody inside the  contract.

A further important feature of the contract is Article 3 that defines specific constraints 
about the off-line behaviour of {\tt Licensor} and {\tt Licensee}, that is their behaviour in the physical world. 
This exemplifies the very general situations where contract's violations cannot 
be fully monitored by the (on-line) software, \emph{i.e.} by the platform that runs the software (either a blockchain or a centralized application), such as the publication of a post in 
a social network, or the leakage of a secret password, %or the violation of the obligation of diligent storage and care, 
or any non-automatically verifiable contextual circumstance.
 The intrinsic \emph{open nature} of legal contracts is exactly this mix of external behaviour and automatic enforcement of contract clauses by means of software. 
%In all these cases, it is required a trusted third party, say an {\tt Authority}, to supervise the disputes occurring from the external monitoring and to provide a trusted dispute resolution mechanism.
The code in Listing~\ref{DigitalLicence} illustrates the \Stipula\ programming pattern that relies on a trusted third party, the {\tt Authority} included in the agreement, to supervise the disputes occurring from the off-line monitoring and to provide a trusted on-line dispute resolution mechanism.

The \texttt{agreement} of Listing~\ref{DigitalLicence} involves three parties: 
\texttt{Licensor} and \texttt{Licensee}, which agree to the parameters of the contract, according to Article 1. (line 6), 
 and \texttt{Authority}, which does not 
need to agree upon the contracts' parameters, but it is important that it is involved 
in the agreement synchronization.
%, because it plays the role of the trusted third party that is entitled to call the functions \texttt{compen}\texttt{sateLicensor} and \texttt{compensateLicensee}.
%
%We say in this case that the Authority party is an \emph{Implicitly} trusted party, since
%in general the parties involved in the agreement need not to know nor to trust each other, but 
%here it is implicitly assumed that both the licensor and the licensee agree to trust the 
%Authority. For instance, \texttt{Authority} could be the address of an online dispute 
%resolution service. We will see in a following example (the \emph{alea} example) the case where 
%the partied need not explicitly agree on the address of a trusted third party. 
By calling the function \texttt{offerLicence}, the {\tt Licensor} transfers to the contract the token corresponding to the full access to the digital service. This transfer is necessary to implement the fact that, after the activation of the the licence (within the agreed time limit {\tt t\_start}, see the event in line 11), the licensor has the legal prohibition of preventing the access to the digital service. 
The {\tt Licensee} can then call \texttt{activateLicence} together with an amount of assets equal to the fixed 
{\tt cost} of 
the license, that is then stored in the {\tt wallet} (line 16). 
In line 17 a fraction of asset is moved towards the authority as a fee, while in line 18 
a personal usage code associated to the token is communicated to the {\tt Licensee}. 

Once entered in the \texttt{Trial} state, the contract can terminate in three ways: 
(\emph{i}) the licensee expresses its willingness to buy the licence by calling 
the function \texttt{buy} which grants him the full token, or (\emph{ii})
 the time limit for the free evaluation period is reached, thus the event scheduled in line 19 
refunds the licensee (but for the fees) and gives the token back to the licensor, or 
(\emph{iii}) during the evaluation period a violation to Article 3 is identified and the authority pre-empts the license by calling either the function \texttt{compensateLicensor} or \texttt{compensateLicensee}. Notice that it is important that the code guarantees that in all possible cases the assets, both the token and the wallet, are not indefinitely locked in the contract.

\subsection{Bike Rental contract: access to a good without transfer of ownership}

We now consider a realistic contract for a city bike rental service\footnote{For instance see the contract in {\scriptsize\tt 
 http://www.thebicyclecellar.com/wp-content/uploads/2013/10/Bike-Rental- Contract-BW.pdf}}, which exemplifies a general rental contract (this is taken from \cite{Festschrift}). 
It involves two parties, the lender and the borrower, which initially agree about what good is rented, 
what use should be made of it, the time limit (or in which case it must be returned), 
the estimated of value and any defects in the good. 
Upon agreement, the payment triggers the legal bond, that is the 
borrower has the permission to use the bike and the lender has the prohibition 
of preventing him from doing so. Note that there is no transfer of ownership, 
but only the right to use the good. 
The contract terminates either when the borrower returns the bike, or when the 
time limit is reached. Litigations could arise when the borrower violates the 
obligations of diligent storage and care, the obligations of using the good only as intended, and not granting the use to a third party without the lender's consent. In these cases the lender may demand %the immediate return of the object, in addition to 
a compensation for the damage. On the other hand, the borrower is entitled to compensation if the good has defects that were known to the lender but that he did not initially disclose.
 
This example puts forward the fact that, when a legal contract refers to a \emph{physical} good, the digital contract needs a digital handle (an avatar) for that good. Moreover, the rent legal contract grants just the \emph{usage} of a good without the transfer of ownership. Many technological solutions, such as smart locks of IoT devices, are actually available. In {\smallang} we abstract from the specific nature of such a digital handle, and we simply represent it as an asset, which intuitively corresponds to a non fungible token associated to the physical good. Moreover, while the communication of the token provides full control of the associated physical good, we assume an operation \texttt{uses(token)} (resp. \texttt{use\_once(token)} or \texttt{uses(token,A)}) that generates a usage-code, say a string, providing access to the object associated to the token (resp. a usage-code only valid (once) for the party \texttt{A}). Therefore, a physical object can be handled as a digital one using the same pattern used in  
the digital license contract above.

\begin{figure}[p]
\vspace*{-2cm}%\hspace*{-3cm}
\includegraphics[scale=1]{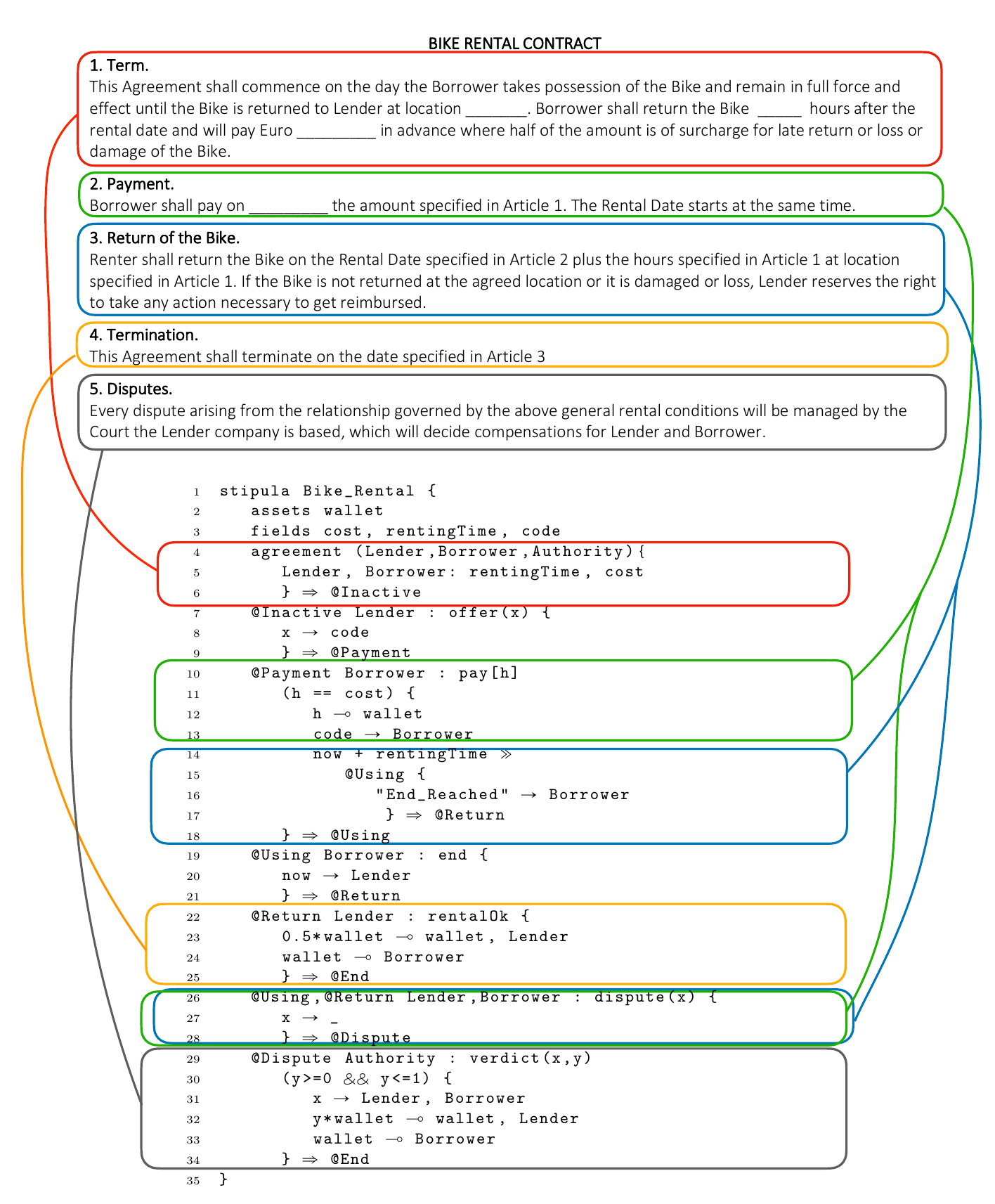} 
\caption{\label{fig:twoversions} A standard Bike Rental contract and its modelling
in {\Stipula}}
\end{figure}
Figure~\ref{fig:twoversions} uses connected boxes to highlight the correspondence between the normative elements of a standard bike rental contract and the corresponding editing in {\Stipula}. 
The parties agrees on the time limit for the rental and the cost of the service, which corresponds to the double of the fee in order to safeguard lender from damages, late returns or loss of the bike. 
For simplicity, in this code the {\tt Lender} sends to the contract a simple usage code for the bike by calling the function {\tt offer}.
Then the {\tt Borrower} pays the expected amount and receives the bike's usage code. Lines 14-18 issue an event corresponding to the obligation of returning the bike within the agreed time limit. %ßMore precisely, 
Indeed, at time {\tt now + rentingTime} the event is automatically triggered by the systems, and if the bike has not been already returned (\emph{i.e.}, the state of the contract is still {\tt @Using}), a message of returning the bike is sent to the borrower and the contract moves to the state {\tt @Return}.
The termination of the rental requires the {\tt Borrower} to call the function {\tt end}, after which the {\tt Lender} has to confirm the absence of damages by invoking {\tt rentalOK}. Only this sequence of actions allows the lender to be payed and the borrower to get back the money deposited as security. For the sake of simplicity this contract does not impose a penalty to the borrower for late return, but it is not difficult to modify the code with an additional state {\tt @LateReturn} so to let the {\tt Lender} keep the entire contract's wallet when {\tt rentalOK} is called in the state {\tt @LateReturn}.

The function {\tt dispute} may be invoked either by the {\tt Lender} or by the 
{\tt Borrower}, either in state {\tt @Using} or {\tt @Return},
and carries the reasons for kicking the dispute off ({\tt x} is intended to
be a string). Once the reasons are communicated to every party (we use the abbreviation
``$\zero$'' instead of writing three times the sending operation)
the contract transits
into a state {\tt @Dispute} where the {\tt Authority} will analyze the issue
and emit a verdict. This is performed by permitting in the state {\tt @Dispute} only the invocation of the {\tt verdict} function, that has two
arguments: a string of motivations {\tt x}, and a coefficient {\tt y} that 
denotes the part of the wallet that will be delivered to {\tt Lender}  
as reimbursement; the {\tt Borrower} will get the remaining part.
It is worth to spot this point: the statement {\tt y*wallet $\lolli$ wallet, Lender}
\emph{takes} the {\tt y} part of {\tt wallet} ({\tt y} is in [0..1])
and sends it to {\tt Lender}; \emph{at the same time} the
{\tt wallet} is reduced correspondingly. The remaining part
is sent to {\tt Borrower} with the statement
{\tt wallet $\lolli$ Borrower} (which is actually a shortening for
{\tt 1*wallet $\lolli$ wallet, Borrower}) and the {\tt wallet} is emptied.

\subsection{Bet contract: dependency on external data}

The bet contract is a simple example of a legal contract that contains an element 
of randomness (\emph{alea}), \emph{i.e.}~where the existence of the performances 
or their extent depends on an event which is entirely independent of the will of 
the parties.
The main element of the contract is a future, aleatory event, such as the 
winner of a football match, the delay of a flight, the future value of a company's stock.

A digital encoding of a bet contract requires that the parties explicitly 
agree on the source of data, % that will determine the final value of the aleatory event 
usually an accredited web page or a specific online service – stored in the field {\tt data_source} – that will publish the final value of the aleatory event. This value will be communicated by the party that assumes the role of {\tt DataProvider}, taking the legal responsibility of supplying the correct data from the agreed source. In particular, it is not necessary that the actual data is directly provided by a trusted institution or an accredited online service, such as an Oracle service, who could hardly take an active legal responsibility in a bet contract. But two betters, say Alice and Bob, can agree to rely on a third party Carl for supplying data, or they can simply agree on the fact that Alice takes both the role of {\tt Better1} and {\tt DataProvider}. 

It is also important that the digital contract provides precise time limits 
for accepting payments and for providing the actual value of the aleatory event. 
Indeed there can be a number of issues: the legal bond must be established before the occurrence of the aleatory event, the aleatory event might not happen, 
\emph{e.g.}~the football match is cancelled, or the data provider might fail to provide 
the required value, \emph{e.g.}~the online service is down.

\begin{lstfloat}
{\footnotesize
\begin{lstlisting}[numbers=left,,numberstyle=\tiny,mathescape,basicstyle=\ttfamily,,caption={The contract for a bet},captionpos=b,label=BetContract] 
stipula Bet {
  assets wallet1, wallet2
  fields alea_fact, val1, val2, data_source, fee, amount, t_before, t_after

  agreement(Better1,Better2,DataProvider){
    DataProvider, Better1, Better2 : fee, data_source, t_after, alea_fact
    Better1, Better2 : amount, t_before                  
   } $\tostate$ @Init
    
  @Init Better1 : place_bet(x)[h]
    (h == amount){ 
        h $\lolli$ wallet1
        x $\send$ val1
        t_before $\eventD$ @First { wallet1 $\lolli$ Better1 } $\tostate$ @Fail
    } $\tostate$ @First

  @First Better2: place_bet(x)[h]
    (h == amount){ 
        h $\lolli$ wallet2
        x $\send$ val2  
        t_after $\eventD$ @Run {
            wallet1 $\lolli$ Better1
            wallet2 $\lolli$ Better2 } $\tostate$ @Fail
  } $\tostate$ @Run
    
  @Run DataProvider : data(x,y,z)[]
    (x == data_source $\&\&$ y==alea_fact){ 
        if (z==val1 $\&\&$ z != val2){              // The winner is Better1             
            fee $\lolli$ wallet2,DataProvider
            wallet2 $\lolli$ Better1
            wallet1 $\lolli$ Better1
        }     
        else if (z==val2 $\&\&$ z != val1){         // The winner is Better2             
            fee $\lolli$ wallet1,DataProvider
            wallet1 $\lolli$ Better2
            wallet2 $\lolli$ Better2
        }
        else {                                     //No winner 		    
            fee*0.5 $\lolli$ wallet1,DataProvider
            fee*0.5 $\lolli$ wallet2,DataProvider
            wallet2 $\lolli$ Better1
            wallet1 $\lolli$ Better1

        } 
   } $\tostate$ @End   
} 
\end{lstlisting}
}
\end{lstfloat}

The {\smallang} code in Listing~\ref{BetContract} corresponds to the case where \texttt{Better1} and
\texttt{Better2} place in \texttt{val1} and \texttt{val2} their bets, while the agreed amount of currency is stored in the contract’s assets {\tt wallet1} and {\tt wallet2}\footnote{For simplicity, this code requires {\tt Better1} 
to place its bet before {\tt Better2}, however it is easy to add similar function to let
the two bets be placed in any order.}. Observe that both bets must be placed within 
an (agreed) time limit \texttt{t\_before} (line 14), to ensure that the legal bond 
is established before the occurrence of the aleatory event. 
The second timeout, scheduled in line 21, is used to ensure the contract termination 
even if the \texttt{DataProvider} fails to provide the expected data, through the 
call of the function \texttt{data}. 

Compared to the Authority pattern in the Digital Licence and Bike Rental examples, the role of the 
\texttt{DataProvider} here is less pivotal than that of the \texttt{Authority}. 
While it is expected that {\tt Authority} will play its part, {\tt DataProvider} 
is much less than a peer of the contract, that is entitled (and legally bound) to call the contract’s function to supply the expected external data. The crucial point of trust here is the 
\texttt{data\_source}, not the \texttt{DataProvider}. As usual, any dispute that might render the contract voidable or invalid, \emph{e.g.}, one better knew the result of the match in advance, or the DataProvider supplied an incorrect value, can be handled by including an {\tt Authority} party, according to the pattern illustrated above.

\subsection{Safe Remote Purchase contract: a distributed interaction protocol}

In a remote purchase\footnote{This example is taken from {\tt
https://docs.soliditylang.org/en/develop/solidity-by-example.html \#safe-remote-purchase}, but most of e-commerce platforms has similar use cases.},
the buyer would like to receive an item from the seller and the seller would like to get money (or an equivalent) in return. The problematic part is the shipment: there is no way to determine for sure that the item arrived at the buyer. The typical solution is to define the interaction protocol so that both parties have \emph{an incentive to resolve the situation} or otherwise their money is locked forever.

\begin{figure}[t]
\begin{center}
\includegraphics[width=6cm]{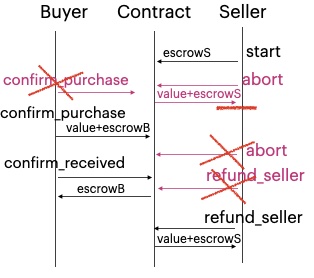}
\end{center}
\caption{Safe Remote Purchase}
\label{fig:purchase}
\end{figure}

The idea is that both parties have to put an amount into the contract as escrow. As soon as this happened, the money will stay locked inside the contract until the buyer confirms that he received the item. The intended protocol is the following sequence of actions (depicted in black in Figure~\ref{fig:purchase}): 
(1) the seller starts the transaction sending its escrow to the contract, (2) the buyer confirms the purchase by sending to the contract the money corresponding to the price of the good plus the escrow, (3) upon reception of the good, the buyer has to confirm the reception to the contract in order to get back the escrow, (4) finally the seller can receive from the contract the  price of the good and the money he deposited in escrow.

Besides the intended sequence of actions, many situations can happen in a remote purchase:
\begin{itemize}
\item if the seller starts but the buyer does not confirm the purchase, seller can take back its escrow with a call to {\tt abort},
\item if the seller does not start, the buyer does not send the escrow, so no money is locked,
\item if the buyer confirms the purchase, the seller cannot take back its escrow (and the payment) until he sends the good (and it is received),
\item if the buyer has confirmed the purchase but he does not confirm the reception, either because the good is not arrived or because the Buyer is  cheating, nobody can take back escrow. Therefore we add to the contract a time limit, after which it is up to an Authority party to decide off-line who is to blame and then implement the decision by calling {\tt refund}. In other terms, the mutual escrow is used as an incentive for the parties to collaborate, but \emph{progress is not ensured} thus the contract requires \emph{timeouts}.
\end{itemize}

 \begin{lstfloat}
{\footnotesize
\begin{lstlisting}[numbers=left,numberstyle=\tiny,mathescape,basicstyle=\ttfamily,caption={The safe remote purchase contract},captionpos=b,label=purchase] 
stipula Purchase {
  asset wallet
  field value, escrow

  agreement(Buyer,Seller,Authority){
     Seller, Buyer, Authority, : value, escrow, time_limit
  } => @Init

   @Init Seller : start [h] (h == escrow) { h $\lolli$ wallet } =>@Created

   @Created  Seller: abort { wallet $\lolli$ Seller } =>@Inactive

   @Created Buyer : confirmPurchase [h] (h == value + escrow) {
       h $\lolli$ wallet
       now + time_limit >> @Locked {
                         "nothing received (maybe!)" $\send$ Buyer
                         "nothing received (maybe!)" $\send$ Seller
                       }=> @Dispute
   } =>@Locked

   @Locked Buyer : confirmReceived { escrow $\lolli$ wallet, Buyer  } =>@Release
   
   @Release Seller : refundSeller  {
      wallet $\lolli$ Seller  // equal to (value+escrow) $\lolli$ wallet, Seller
   } =>@Inactive

   @Dispute Authority : refund(x,y) (wallet==x+y) {
       x $\lolli$ wallet, Buyer
       y $\lolli$ wallet, Seller
   } =>@Inactive
}
\end{lstlisting}
}
\end{lstfloat}

We remark that the contract in Listing~\ref{purchase} does not solve all legal issues. For instance in a purchase the consumer has the power to withdraw from an online sale, and there are usually warranties if the good was damaged or different from the sellers' description. To deal with all these situations the contract can be enriched with a more complex Authority pattern as in the previous examples.

%\item A precondition is not met, e.g. abort is not called by Seller, or is called at a bad contract's state, or the parameters have bad values. In this case the transaction is not executed. {\bf DO WE NEED TO ISSUE A REVERT???}
%\item Send a negative asset, e.g.  value - 100 --o A, or (balance --o A ; balance --o B). It is an error, i.e. stuck.
%\end{itemize}

\subsection{Mutual Dissent and contract modification}

There is a last distinctive element in legal contracts that deserves a comment:
the management of exceptional behaviours, that is all those behaviours that cannot be 
anticipated due to the occurrence of unforeseeable and extraordinary events. 
For instance, legal contracts can always be dissolved if the parties agree. 

We can model the \emph{mutual dissent} by including a specific function in the contract, which can be activated with the agreement of both parties, that causes the contract to go into a stand-by state, which blocks the execution of all functions not yet performed. This prevents the contract from continuing when both parties no longer want it to.
More precisely, the following code shows the Mutual Dissent pattern for a generic contract {\tt C} where parties {\tt P1} and {\tt P}2 may express mutual dissent:

\begin{lstfloat}
{\footnotesize
\begin{lstlisting}[mathescape,basicstyle=\ttfamily,caption={The mutual dissent pattern},captionpos=b,label=MutualDissent] 
stipula Rescindable_C {
  assets a1, ..., an
  fields ...
  agreement(P1,P2,....,Authority) ....

  // add a copy of this function for any state X of the contract C
  @X P1: dissent { now + 1 day >> @OneDissented { }=>@X } =>@OneDissented
  @OneDissented P2: dissent {} => @Rescinded
  
  @X P2: dissent { now + 1 day >> @TwoDissented { }=>@X } =>@TwoDissented 
  @TwoDissented P1: dissent{} => @Rescinded
  
  @Rescinded  Authority : terminate {
    a1  --o Authority
    ...
    an  --o Authority
  } =>@End
\end{lstlisting}
}
\end{lstfloat}

To prevent assets being locked indefinitely in the contract, the function {\tt terminate} sends all the assets to the authority. More complex assets reallocation to the parties can also be implemented, provided that they mutually agree on the reallocation. % This would require a function allowing one party to specify where the assets go to what extent and the other to accept the distribution. ATTENZIONE: i nomi dei parties non possono essere parametri attuali!
%\begin{verbatim}
% @Rescinded  Authority : terminate (x1,...,xn) {
%    a1  --o x1
%    ...
%    an  --o xn
%  } ==> @End
%\end{verbatim}
%
%{\bf Però questa suddivisone deve essere prevista alla sottoscrizione iniziale del contratto}, per decidere dinamicamente come suddividere gli asset serve cambiare codice dinamicamente, cioè cambiare contratto.

%A simpler pattern can be used when only one party requests blocking and asset recovery, according to a certain allocation plan. In this case the party can suspend the normal execution (with a call to the following function {\tt block}), and the authority agrees.
%use any new ad-hoc feature, rather a simple pattern is provided that defines a template:
%%% NOOO: così se l'autorità decide di non sospendere, non si sa a che stato ritornare!
%{\footnotesize
%\begin{lstlisting}[mathescape,basicstyle=\ttfamily] 
%  ~@End _ : block(x) {
%     x $\send$ _
%     } $\tostate$ @Exception
%
%  @Exception Authority : handle(x,y) //similar to verdict(x,y) 
%\end{lstlisting}}
%
%
%According to the above pattern, the function {\tt block} may be invoked by any party
%(notation ``$\zero$'') provided the lifetime of the 
%contract is \emph{not terminated} (the contract is not in the state {\tt End}).
%The management of the exception is
%similar to that of disputes and therefore omitted.
%

Finally, parties have the power of dynamically change the terms of the contract if they agree to it. 
Contract modification can be modelled by the termination of the running contract {\tt C} (with the mutual dissent pattern), and the activation of a new contract {\tt C'}, to which the assets remaining in {\tt C} are transferred.
The basic \Stipula\ language does not allow to pass contracts' names as arguments, nor allows to invoke external contracts' activations or inter-contracts functions invocations (differently from, \emph{e.g.} Solidity smart contracts). Therefore, the bridge between the termination of {\tt C} and the activation of {\tt C'} with remaining assets must be performed off-line by the Authority.

\section{Legal contracts and the power of formal methods}

As already discussed, the advantage of using a Legal Calculus to draft legal contracts is that a concise and well-defined language reduces the ambiguities (and therefore the grey areas) characteristics of traditional legal drafting. In particular, we remark that there are three levels of formalisation, corresponding to 
%highlight the different merits coming form the formalisation of 
three different aspects of a language: the syntax, the semantics, and the analysis and verification tools.

Almost all projects for Code-Driven Law put forward a legal language based on a well-defined syntax. This is indeed the base to mechanise the writing of dedicated software that encodes legal content --not just legal contracts but any kind of legal data. These projects often come with templates for standard legal documents, that can be customised by setting template’s parameters with appropriate values. There are legal language definitions based on context free grammars (as Lexon~\cite{Lexon}), or domain specific markup languages and ontologies to wrap logic and
other contextual informations around traditional legal prose (as OpenLaw~\cite{OpenLaw}, Accord~\cite{Accord}, SLCML~\cite{SLCML}), or legal specification languages based on visual programming interfaces (as in \cite{Babbage,Blockly}).

More complex is instead the formalisation of the semantics of a programming language, which is however essential to have the full understanding of the software, and the certainty of the dynamic contract's behaviour. Legal calculi, such as Catala~\cite{Catala}, Orlando~\cite{Orlando} and Stipula~\cite{techReport}, have the suitable size to fully handle their formal semantics. We discuss below the case of {\Stipula}, which acknowledges the concurrent nature of legal contracts as interaction protocols, and resorts to concurrency theory to define the semantics of contracts and to precisely control complex aspects like nondeterminism.

Finally, the most powerful benefit of formal methods is the deployment of automated tools to (statically) analyse the legal software in order to check safety properties, verify the absence of specific errors and possibly the reachability of convenient states. This level of formalisation is still at an initial stage in the literature, since it requires a robust definition (and implementation) of the language semantics, and because the identification of the desirable properties of legal software is still an open question.

%requires well-defined categories and the precise stipulation of methods and conditions that need to be defined in advance. 
%
%the ambiguity of human language and the need for legal norms to be flexible and fact dependent.
%
% As opposed to legal rules, written as general rules in a natural language that is inherently ambiguous, technical rules can only be implemented into code, and thus necessarily rely on formal algorithms and mathematical models. Regulation by code is therefore always more specific and less flexible than the legal provisions it purports to implement.
% 
% . In the long run, the lack of textual ambiguity might reduce the need for canons of construction and other textual interpretation techniques — although factual ambiguity (i.e., did a real world event happen or not) will obviously remain.
% 

\subsection{Defining a formal semantics}
The full definition of \Stipula's operational semantics (currently submitted to publication, but a preliminary version is available in \cite{techReport}) is given in terms of a labelled transition system 
$\contract {\comma} \Time \lred{\mu} \contract' {\comma} \Time'$ that highlights the open nature of the contracts' behaviour, whose execution requires the interaction with the external context. The \emph{runtime configuration} $\contract {\comma} \Time$ is a pair where $\contract$ is the runtime status of the running contract (storing its current state and the pending events), and $\Time$ is the time value of the system's global clock. The actions that can be performed by a contract time $\Time$ are the following
$$ \mu \quad ::= \quad \tau  \quad |\quad  (\vect A,\, \vect{A_i} :  \vect{v_i}^{i\in 1,..,n}) \quad |\quad  
                 A : \texttt{f}(\wt{u})[\wt{v}] \quad |\quad
                 v \send A \quad |\quad  a \lolli A \; .
$$
where the label $(\vect A,\, \vect{A_i} :  \vect{v_i}^{i\in 1,..,n})$ observes the agreement that the parties are going to sign, that is who is taking the legal responsibility for which contract’s role, and what are the terms of the contract, \emph{i.e.}, the agreed initial values of the contract’s fields. The label $A : \texttt{f}(\wt{u})[\wt{v}]$ observes the possibility (at time $\Time$) for the party $A$ to call the function $f$. The labels $v \send A$ and $a \lolli A$ observe  that (at time $\Time$) the party $A$ can receive a value and an asset, respectively. Contract's field updates, internal asset moves and event scheduling, as well as time progress, are not observed (label $\tau$).

\noindent
The behaviour of a {\Stipula} legal contract can be described as the following procedure: 
\begin{enumerate}
\item 
the first action is always an agreement, which moves the contract to an idle state;
\item 
in an idle state, if there is a ready event with a matching state, then its handler is 
completely executed, moving again to a (possibly different) idle state;
\item 
in an idle state, if there is no event to be triggered, either advance the system's clock or
call any permitted function (\emph{i.e.}~with matching state and preconditions). A function
invocation amounts to execute its body until the end, which is again an idle state.
\end{enumerate}
Therefore, the semantics has three sources of nondeterminism: 
(\emph{i}) the order of the execution of ready events' handlers, 
(\emph{ii}) the order of the calls of permitted functions, and (\emph{iii}) the delay of 
permitted function calls to a later time (thus, possibly, after other event handlers). 
We remark that a nondeterministic behaviour is not necessarily an error: even the execution of legal contracts written in natural language might lead to nondeterministic executions, in particular when the contract leaves room for a participant  not to timely perform an action that was expected to do. Depending on how the contract is written, this may be admissible or may cause a legally uncertain situation that can only be solved by a court. Therefore, the precise formalisation of a contract’s behaviour in terms of an operational semantics has the advantage of explicitly knowing what are the sources of nondeterminism, and allows to precisely control it.

\subsection{Observing legal contracts through Normative Equivalence}

One of the difficulties of writing contracts in natural language is the fact that the same legal bindings can be expressed with many similar texts. Then it is often difficult to properly check when two contracts that are syntactically different are instead legally equivalent, meaning that the parties using them cannot distinguish one from the other.
By relying on the operational semantics, that formally defines the observable actions of a contract behaviour, we can define a \emph{bisimulation-based observation equivalence}, where two contracts are deemed to be legally equivalent if they involve the same parties observing the same interactions during the contracts' lifetime. 

More precisely, the so-called \emph{Normative Equivalence} (see \cite{techReport}) 
%is defined in terms of largest bisimulation $\mathcal{R}$ between runtime configurations such that whenever
%$\contract_1 {\comma} \Time \ \ \mathcal{R}\ \ \contract_2 {\comma} \Time$
%\begin{enumerate}
%\item
%if $\contract_1 {\comma} \Time \Lred{(\vect A,\, \vect{A_i} :  \vect{v_i})} \contract_1' {\comma} \Time$ then 
%% there exists $\alpha'$ with $\alpha\sim\alpha'$ such that
%$\contract_2 {\comma} \Time \Lred{(\vect A,\, \vect{A_i} :  \vect{v_i})} \contract_2' {\comma} \Time$ and 
%$\contract_1' {\comma} \Time \ \ \mathcal{R} \ \ \contract_2', \Time$;
%\item
%if $\contract_1 {\comma} \Time \Lred{\mu_1} \cdots \Lred{\mu_n} \contract_1', \Time \lred{}\contract_1', \Time+1$
%then there exist $\mu_1' \cdots \mu_n'$ that is a permutation of $\mu_1 \cdots \mu_n$ 
%such that
%$\contract_2 {\comma} \Time \Lred{\mu_1'} \cdots \Lred{\mu_n'} \contract_2'{\comma}\Time
% \lred{} \contract_2'{\comma} \Time+1$  and $\contract_1' {\comma} \Time+1 \ \mathcal{R} \ \contract_2' {\comma} \Time+1$.
%\end{enumerate}
%\noindent
equates two contracts if %, at the same time unit $\Time$
\begin{itemize}%\itemsep-1pt
\item they provide the same agreement, that is the same parties take the same legal responsibility  and agree on the same terms of the contract (expressed by the action $(\vect A,\, \vect{A_i} :  \vect{v_i}^{i\in 1..n})$); 
\item 
every party is subject to  the same dynamic set of permissions, prohibitions and obligations; 
\item every party receives from the contract the same assets and values (actions $v \send A$ and $a \lolli A$);
\item the bisimulation game abstracts away the ordering of the observations within the same time clock, and enforces a transfer property that shifts the time of observation to the next time unit.
\end{itemize}
To observe the permissions and prohibitions at time $\Time$, we observe whether any party can invoke, resp. cannot invoke, any function (expressed by the action $A : \texttt{f}(\wt{u})[\wt{v}]$). Obligations are captured implicitly by shifting the observation at a specific point in time, and observing –in the future– the effects of executing the event that encodes the legal commitment.
In particular, the system’s clock needs not to be directly observed: by checking the set of permissions and prohibitions at any time units, and since only a contract’s state change can modify the set of valid permissions and prohibitions, it is sufficient to observe whether a function can be executed before of after another function or an event, disregarding its precise execution time unit.

%Moreover, the contract equivalence studied in the next section shows that time is reasonably observed only through obligation deadlineTherefore, in this paper we have preferred to stick to the simpler semantics and accept a fair implementation.
As a consequence, the Normative Equivalence safely abstracts away the ordering of the observations within the same time unit: if a party receives two messages in different order it might be due to delays of communications, rather to sensible differences in the contracts. Nevertheless, the equivalence does not overlook essential precedence constraints, which are important in legal contracts, as the requirement that a function delivering a service can only be invoked after another specific function, say a payment. Additionally, the Normative Equivalence abstracts away from the names of the contract’s assets, fields and internal states, and it is also independent from future clock values, allowing to garbage-collect events that cannot be triggered anymore because the time for their scheduling is already elapsed. %Finally, a set of algebraic laws formalize that the ordering of communications can be safely overlooked, as long as they belong to the same global time.

\subsection{Verification of contracts' properties}

By looking at legal contracts as interaction protocols and by relying on a well defined operational semantics, 
the rich theory of formal methods for concurrent systems can be a great source of inspiration to develop analysis and verification tools. However, first of all it is essential to conduct an interdisciplinary investigation to properly identify what are the errors and the properties that should be targeted by the techniques providing safety and liveness guarantees.

An important class of errors are those related to unsafe usage of assets, which must obey to a linear semantics (no forging, no duplication, no loss) and whose content must be meaningful. For instance, in \Stipula\ the assets corresponding to currency, as the asset {\tt wallet} in the examples above, must always contain a non negative amount of money.  Accordingly, an asset transfer what would leave a contract's asset with a negative (unsafe) asset, \emph{e.g.},  {\tt 100 $\lolli$ wallet,A} when {\tt wallet} holds less than 100 coins, is not executed and results in a stuck configuration. Similarly, if the contract's asset {\tt token} already contains a non fungible token providing access to a good, say a digital service, then the operation {\tt t $\lolli$ token} that would accumulate or overwrite the {\tt token} with the asset {\tt t} must not be executed. Moreover, assets must not be indefinitely locked into contracts: at any time it should be possible, at least for some party, to redeem the assets stored into the contract; this is often called \emph{liquidity} property (\cite{BartZ2019}). 
These issues are at the core of the research about \emph{resource-aware languages} as Obsidian~\cite{Obsidian,ObsidianUsage}
Nomos~\cite{Nomos,Resources20}, Flint~\cite{Flint} and Move~\cite{Move}; and even the questions "What is the type {\tt Money} in a programming language? What are its suitable abstractions?" and "What is the difference between the more general type {\tt Asset} and the type {\tt Money}?" are still open issues. 
%
% In fact, the above languages provide type systems that guarantee that assets are not 
%accidentally lost, even if none of them address liquidity.
%More precisely, Obsidian uses types to ensure that owning references to assets cannot be lost unless they are explicitly disowned by the programmer. Nomos uses a linear type system to prevent the duplication or deletion of assets and amortized resource analysis to statically infer the resource cost of transactions. 
%%
%Finally, Marlowe~\cite{Marlowe}, being a language for financial contracts,
%does not admit that money be locked forever in a contract. In particular, Marlowe's 
%contracts have a finite lifetime and, at the end of the lifetime, any remaining money 
%is returned to the participants. In other terms, all contracts are liquid by construction. 
%In the extension of {\Stipula} with events, the finite lifetime constraint 
%can be explicitly programmed: a contract issues an event at the beginning 
%so that at the timeout all the contract's assets are sent to the parties. 

Other kinds or errors are those related to non collaborative parties, that might prevent the progress of the contract or might move it to a problematic state. We have described \Stipula\ design patterns, as the authority pattern or the mutual dissent pattern, that can be inserted in the drafting of the digital contract as a sort of escape hatch; however, a static analysis of the runtime behaviour of the contract would be very useful.

\section{Conclusions} %Open issues for further work}

In this article we discussed the role of Legal Calculi in the process of digitisation of legal contracts.
We illustrated the design choices of \Stipula, whose primitives naturally support the encoding of contracts’ normative elements (permissions, prohibitions, obligations, asset transfer, judicial enforcement and openness to the external context). We also remarked that legal contracts can be interpreted as interaction protocols between concurrent parties, leading to a fruitful connection with the rich toolset of formal methods available for concurrent systems.

Studying the theory of domain-specific legal calculi is a useful research line, that supplement the development of the Code-Driven Law trend. On the other hand, it is important to keep a lively connection between these calculi and other two fundamental abstraction levels: the effective implementation and the interdisciplinary assessment.
The actual implementation of legal calculi brings in specific challenges, such as the \emph{legally robust} management of the identities of the parties and their valid agreement to the legal bonds. Moreover, the implementation of obligations by scheduling an event that issues a corresponding penalty if the obligation has not been met, may not be always feasible, and asks for an accurate management of time, which is a well-known challenge in distributed platforms.

Finally, the dialogue with legal researchers and professionals provides valuable insights, non just on the usability of legal programming languages, but mainly on the actual meaning (in the epistemic sense) of their abstractions. This is important to unveil when partial or erroneous interpretations of the law has been embedded in the technical artefacts, and to understand the actual extent of the legal protection provided by the software normativity.
A main lesson that we learned is the intrinsic open nature of legal contracts, that is incompatible with the automatic execution of software-based rules claimed by the Code-Driven Law. Indeed, a contract produces the intended effects, declared by the parties, only if it is legally valid: the law may deny validity to certain clauses, as an excessive interests rate. The intervention of the law is particularly significant when the contractor (usually the weaker party, such as the worker in an employment contract or the consumer in an online purchase) agrees without having awareness of all clauses in the contract, nor having the ability to negotiate them, due to the existing unbalance of power (\cite{techReport}). Therefore, any technical solution based on a legal programming language
must provide an escape mechanism  (as the authority pattern in \Stipula) that allows a flexible, and legally valid, link between what is true on-line and off-line.

%\nocite{*}
\bibliographystyle{eptcs}
\bibliography{bibliography}
\end{document}